# Self-Adaptive Spike-Time-Dependent Plasticity of Metal-Oxide Memristors

M. Prezioso[1], F. Merrikh-Bayat[1], B. Hoskins[1], K. Likharev[2], and D. Strukov[1]

## Abstract

Metal-oxide memristors have emerged as promising candidates for hardware implementation of artificial synapses – the key components of high-performance, analog neuromorphic networks - due to their excellent scaling prospects. Since some advanced cognitive tasks require spiking neuromorphic networks, which explicitly model individual neural pulses ("spikes") in biological neural systems, it is crucial for memristive synapses to support the spike-time-dependent plasticity (STDP), which is believed to be the primary mechanism of Hebbian adaptation. A major challenge for the STDP implementation is that, in contrast to some simplistic models of the plasticity, the elementary change of a synaptic weight in an artificial hardware synapse depends not only on the pre-synaptic and post-synaptic signals, but also on the initial weight (memristor's conductance) value. Here we experimentally demonstrate, for the first time, STDP protocols that ensure self-adaptation of the average memristor conductance, making the plasticity stable, i.e. insensitive to the initial state of the devices. The experiments have been carried out with 200-nm $Al_2O_3/TiO_{2-x}$ memristors integrated into 12×12 crossbars. The experimentally observed self-adaptive STDP behavior has been complemented with numerical modeling of weight dynamics in a simple system with a leaky-integrate-and-fire neuron with a random spike-train input, using a compact model of memristor plasticity, fitted for quantitatively correct description of our memristors.

[1] Department of Electrical and Computer Engineering, University of California at Santa Barbara, Santa Barbara, CA 93106. [2] Department of Physics and Astronomy, Stony Brook University, Stony Brook, NY 11794.

## Introduction

In biological neural systems, neurons communicate with each other with action potential pulses - "neural spikes".[1] While some of the network activity information is encoded in the average spiking rate, many experiments in neurobiology suggest that the timing of individual spikes matters, and is essential for coordinated processing of temporal and spatial information.[2,3] Indeed, encoding information with single spikes or inter-spike intervals provides a higher information capacity than the firing-rate codes which represent only the average spiking activity.[4] These observations motivated the development of spiking neuromorphic hardware circuits which explicitly model neural spikes.[1,4] An additional motivation[5-7] for pursuing spiking neuromorphic networks is their higher energy efficiency, recently demonstrated in a very large system.[8]

In the simplest spiking neuromorphic networks, each neuron is modeled as a leaky-integrate-and-fire unit, which integrates incoming spikes and fires its own spike when the integrated action potential reaches a certain threshold.[1] The fired spike, weighed according to the strengths of the corresponding synapses, is applied to the input of other neurons. Additionally, the fired spike is also propagated backwards to the input synapses to provide their weights' adaptation – for example, according to the spike-timing-dependent-plasticity (STDP) rule,[9-12] which ensures Hebb-like learning.[9] For the most common STDP type, found for example in Layer 5 of the neocortex,[13] the synaptic weight is increased if the post-synaptic spike follows soon after the pre-synaptic one (implying their causal relation), is decreased if their timing order is opposite (implying a random coincidence of the spikes), and is virtually unaltered if the time interval $\Delta t$ between the spikes is larger than a few milliseconds.

The STDP-enabling hardware based on traditional integrated circuit technologies, in which synaptic weight values are stored, for example, in digital static-random access memories,[8] or as analog charges in switched capacitor structures,[7] can hardly ensure the network density necessary for cortex-scale systems. On the other hand, the values may be stored as conductivities of very compact, two-terminal, nonvolatile devices, "memristors", which may be scaled down to ensure such density.[14,15,16] This is why, following several suggestions of STDP implementation in memristors, using various shapes of pre- and post-synaptic pulses,[17,18,19,20,21,22] there has been a recent surge of experimental demonstrations of the STDP functionality in organic,[23,24,25,26] complex-oxide,[27] sulfide,[28,29] silicon-oxide[30], hafnium-oxide[31] and phase-change[32] devices.

In this work we describe the first implementation of the STDP in the most scalable, metal-oxide memristors, whose crossbar integration has already been demonstrated.[32] Most importantly, we have shown that in these devices the STDP may be self-adaptive, excluding the need in the continuous adjustment of the average conductance of each device.

## Results

All experiments were carried out with $Pt/Al_2O_3/TiO_{2-x}/Ti/Pt$ memristors integrated in $12\times12$ crossbar circuits (Fig. 1a) – see the Methods section for fabrication details. Fig. 1b shows a typical switching hysteresis of a crossbar-integrated memristor at a quasi-DC symmetric voltage sweep. The ON/OFF current ratio measured at a non-disturbing bias of 0.2 V is close to 10. The results of a detailed electrical characterization of these crossbar-integrated devices, including their switching endurance of at least 5,000 cycles, projected retention time is excess of 10 years, and low variability of forming and switching voltages, were reported earlier.[33]

In the first set of experiments, we have implemented three different biologically-plausible "STDP windows", i.e. the dependences of the synaptic weight change on the time interval $\Delta t$ between the pre- and post-synaptic spikes (Fig. 2). In particular, getting each experimental point shown on the bottom panels (g, h, i) of Fig. 2 involved three steps. First, memristor's conductance $G$, which represents its synaptic weight and was measured at 0.2 V, was set to an initial value $G_0$ $\approx 33$ µS, using a simple but efficient tuning algorithm.[34] Pre-synaptic and post-synaptic pulses of the waveforms shown on one of the top panels (a, b, c) of Fig. 2, selected following the recommendations of Ref. 21, were then applied to the top and bottom wires leading to the selected memristor inside the crossbar, with a certain delay $\Delta t$ between the pulses, while the remaining lines of the crossbar were kept grounded. Finally, after the pulse application, the new value of memristor's conductance was measured and its change calculated. The experiment was performed repeated 10 times for each particular $\Delta t$, every time resetting the device to the same initial conductance with 10% accuracy. As Fig. 2 shows, these three different spike shapes result in three different representative STDP window shapes found in various biological synapses.[13] Other window shapes, e.g., those which correspond to $\Delta t$ sign flip (and hence may be used for the anti-Hebbian rule implementation), may be readily obtained by changing switching polarity and/or modifying pulse timings.

The initial conductance of 33 μS, chosen for the described set of experiments, crudely corresponds to the middle of the dynamic range for the considered memristors. In the second set of tests, the experiment with waveforms corresponding to the first STDP window (Fig. 2g) was repeated for several different initial values $G_0$ of conductance, spanning the whole dynamic range of our memristors. As this Fig. 3a shows, very naturally there is no increase in conductance when its initial value is close to its maximum value, and no decrease in conductance in the opposite case, i.e. when $G_0$ is close to device's minimum conductance. Such saturation in the switching dynamics is typical for many types of memristors.[15,16,33,34,35]

This strong dependence of memristor's plasticity on its initial state might cause concerns about the possible need in continuous external tuning of each device, which would make large-scale spiking networks impracticable. To investigate this issue, we have carried out numerical simulation of the STDP adaptation, using an analytical, phenomenological ("compact") model of the experimentally observed conductance change for the particular STDP window shown in Fig. 2g. It has turned out that the change is well described by the following product:

$$\Delta G = \Lambda_t \Lambda_G, \tag{1}$$

$$\Lambda_t \equiv \begin{cases} a^+ \left\{ \tanh\left[ b_t^+ \left( \Delta t + c_t^+ \right) \right] - 1 \right\}, & \text{for } \Delta t > 0, \\ a^- \left\{ \tanh\left[ b_t^- \left( \Delta t + c_t^- \right) \right] + 1 \right\}, & \text{for } \Delta t < 0, \end{cases} \tag{2}$$

$$\Lambda_G \equiv \begin{cases} 1 + \tanh\left[ b_G^+ \left( G_0 + c_G^+ \right) \right], & \text{for } \Delta t > 0, \\ 1 - \tanh\left[ b_G^- \left( G_0 + c_G^- \right) \right], & \text{for } \Delta t < 0, \end{cases} \tag{3}$$

where $a$, $b$, $c$ and $d$ are fitting parameters. (In some publications,[36,37] such functions are called "multiplicative"; note, however, that though at each of the two intervals of $\Delta t$, $\Delta G$ is indeed a product of separate functions of $G_0$ and $\Delta t$, globally it is not, since according to Eq. (3), function $\Lambda_G$ depends not only on $G_0$, but also on $\Delta t$ – via its sign. Due to this reason, the plots of $\Delta G$ as a function of $\Delta t$, shown with the continuous surface in Fig. 3b, are not globally self-similar.) As the dots in Fig. 3 show, this function, with an appropriate choice of the fitting parameters, describes the experimentally observed behavior very reasonably. Moreover, we believe that such $G_0$- dependent STDP behavior may be expected for many types of memristive devices with saturating switching dynamics.[15,16,33,34,35]

Using the STDP model so verified, we have simulated the time evolution of memristor conductances in a simple, generic neuromorphic network with just one soma, described with the leaky-integrate-and-fire model, and 100 input synapses (Fig. 4). The network was fed with similar spikes of the shape shown with black lines in Fig. 2a, with random, independent, Poisson-distributed initiation times with 14 Hz average spiking rate. As Fig. 4d shows, memristor conductances eventually evolve to a stable bell-curve distribution independent of their initial values, with the peak of the distribution centered in the middle of the dynamic range. This is not quite surprising, because our model qualitatively corresponds to the typical STDP behavior observed in biology,[11,38] and also to phenomenological "multiplicative" models that predict similar self-adaptation,[36,37] which is deemed necessary for long-term stability of spiking neural networks. (As illustrated by the bottom panel of Fig. 4d, so-called "additive" STDP models, in which $\Delta G$ is independent of $G_0$,[36] cannot ensure such self-adaptation.)

## Discussion

The demonstrated dependence of the STDP window on the applied voltage waveforms (cf. panels (a-c) and (g-i) of Fig. 2) may be readily explained by taking into account that memristor's conductance changes mostly when the net voltage applied to the device exceeds certain switching threshold voltage – see the dashed lines in Fig. 1b and panels (d-f) of Fig. 2. As the result, the conductance change $\Delta G$ semi-quantitatively follows either the time maximum or the time minimum of the applied waveforms – whichever exceeds the corresponding threshold more. Panels (d-f) show these voltage extremal values for the used waveforms (a-c); their comparison with the corresponding experimental STDP windows shown in panels (g-i) indeed confirms their similarity. Some slight deviations from this correspondence, for example, the time asymmetry of the window shown on panel (i), may be readily explained by the switching dynamics dependence on the conductive state of the memristor.

Another unexpected anomaly of the data is the presence of the (weak and broad) second peak in the distribution of final conductances in the numerical simulation of synaptic self-

adaptation – see Fig. 4. This peak might be suppressed by balancing device's asymmetry by slightly varying parameters of the STDP  – see Fig. S1 and its discussion.

## Summary

We have experimentally demonstrated that $Al_2O_3/TiO_2$-based memristors may be used to implement the spike-time dependent plasticity with STDP window shapes similar to those observed in biological neural systems. By fitting the experimental data with a simple compact model, we have shown that such STDP behavior enables self-adaptation of the synaptic weights to a narrow interval in the middle of their dynamic range, at least in a simple (but very representative) spiking network. These results give every hope for stable operation of future large neuromorphic networks based on such memristors.

### Acknowledgments

This work was supported by the Air Force Office of Scientific Research (AFOSR) under the MURI grant FA9550-12-1-0038, DARPA under Contract No. HR0011-13-C-0051UPSIDE via BAE Systems, and DENSO Corporation, Japan. Useful discussions with L. Sengupta and G. Adam are gratefully acknowledged.

## References

1    Gerstner, W. & Kistler, W. *Spiking Neuron Models* (Cambridge U. Press, New York, NY, 2008).

2    Bialek, W., Reike, F., Ruyter van Steveninck, R.R. & Warland, D. Reading a neural code, *Science* **252**, 1854 – 1857 (1991).

3    Palm, G., Aertsen, A.M. H.J. & Geirstein, G.L. On the significance of correlations among neural spike trains, *Biological Cybernetics* **59**, 1-11 (1988).

4    Maas, W. Networks of spiking neurons: The third generation of neural network models. *Neural Networks* **10**, 1659-1671 (1997).

5    Indeveri, G. *et. al.* Neuromorphic silicon neuron circuits. *Frontiers in Neuroscience* **5** (2011).

6    T. Pfeil *et al.* Six networks on a universal neuromorphic computing substrate. *Frontiers in Neuroscience* **7** (2013).

7    Benjamin, B. V. *et al.* Neurogrid: A mixed-analog-digital multichip system for large-scale neural simulations, *Proceedings of the IEEE* **102**, 699–716 (2014).

8    Merola, P. A. *et al*. A million spiking-neuron integrated circuit with a scalable communication network and interface. *Science* **8**, 668-673 (2014).

9    Hebb, D.O. *The Organization of Behavior* (Wiley & Sons, New York, NY, 1949).

10   Gerstner, W., Ritz, R. & Van Hemmen, J. L. Why spikes? Hebbian learning and retrieval of time-resolved excitation patterns. *Biological cybernetics* **69**, 503-515 (1993).

11   Bi, G.-Q. & Poo, M.-M. Synaptic modifications in cultured hippocampal neurons: dependence on spike timing, synaptic strength, and postsynaptic cell type. *Journal of Neuroscience* **18**, 10464-10472 (1998).

12   Caporale, N. & Dan, Y. Spike timing–dependent plasticity: A Hebbian learning rule. *Annual Review of Neuroscience* **31**, 25-46 (2008).

13   Abbott, L. F. & Nelson, S. B. Synaptic plasticity: Taming the beast. *Nature Neuroscience* **3**, 1178-1183 (2000).

14   Likharev, K. K. CrossNets: Neuromorphic hybrid CMOS/nanoelectronic networks. *Science of Advanced Materials* **3**, 322-331 (2011).

15   Pickett, M. D. *et al.* Switching dynamics in titanium dioxide memristive devices. *Journal of Applied Physics* **106**, 074508 (2009).

16   Yang, J. J., Strukov, D. B. & Stewart, D. R. Memristive devices for computing. *Nature Nanotechnology* **8**, 13-24 (2013).

17   Snider, G. S. Spike-timing-dependent learning in memristive nanodevices. *NanoArch'08*, 85-92 (2008).

18   Likharev, K. K. Hybrid CMOS/nanoelectronic circuits: Opportunities and challenges. J. Nanoel. & Optoel. **3**, 203-230 (2008).

19   Linares-Barranco B. & Serrano-Gotarredona T. Memristance can explain Spike-Time-Dependent Plasticity in neural synapses, *Nature Preceedings* http://precedings.nature.com/documents/3010/ (2009).

20   Afifi, A. , Ayatollahi, A. & Rassi, R. STDP implementation using memristive nanodevice in CMOS-nano neuromorphic networks. *IEICE Electronics Express* **6**, 148-153 (2009).

21   Zamarreño-Ramos, C. *et al.* On spike-timing-dependent-plasticity, memristive devices, and building a self-learning visual cortex. *Frontiers in Neuroscience* **5** (2011).

22   Saighi, S. *et al.* Plasticity in memristive devices for spiking neural networks. *Frontiers in Neuroscience* **9** (2015).

23   Alibart, F. *et al.* An organic nanoparticle transistor behaving as a biological spiking synapse. *Advanced Functional Materials* **20**, 330-337 (2010).

24   Li, S. *et al.* Synaptic plasticity and learning behaviours mimicked through Ag interface movement in an Ag/conducting polymer/Ta memristive system. *Journal of Materials Chemistry C* **1**, 5292-5298 (2013).

25    Subramaniam, A., Cantley, K. D., Bersuker, G., Gilmer, D. & Vogel, E. M. Spike-timing-dependent plasticity using biologically realistic action potentials and low-temperature materials. *IEEE Trans. Nanotechnology* **12**, 450-459 (2013).

26    Zeng, F., Li, S., Yang, J., Pan, F. & Guo, D. Learning processes modulated by the interface effects in a Ti/conducting polymer/Ti resistive switching cell. *RSC Advances* **4**, 14822-14828 (2014).

27    Wang, Z. Q. *et al.* Synaptic learning and memory functions achieved using oxygen ion migration/diffusion in an amorphous InGaZnO memristor. *Advanced Functional Materials* **22**, 2759-2765 (2012).

28    Nayak, A. *et al.* Controlling the synaptic plasticity of a $Cu_2S$ gap-type atomic switch. *Advanced Functional Materials* **22**, 3606-3613 (2012).

29    Ohno, T. *et al.* Short-term plasticity and long-term potentiation mimicked in single inorganic synapses. *Nature Materials* **10**, 591-595 (2011).

30    Jo, S. H. *et al.* Nanoscale memristor device as synapse in neuromorphic systems. *Nano Letters* **10**, 1297-1301 (2010).

31    Kuzum, D., Jeyasingh, R. G., Lee, B. & Wong, H.-S. P. Nanoelectronic programmable synapses based on phase change materials for brain-inspired computing. *Nano Letters* **12**, 2179-2186 (2011).

32    Mandal, S., El-Amin, A., Alexander, K., Rajendran, B. & Jha, R. Novel synaptic memory device for neuromorphic computing. *Nature Scientific Reports* **4**, 5333 (2014).

33    Prezioso, M. *et al.* Training and operation of an integrated neuromorphic network based on metal-oxide memristors. *Nature* **521**, 61-64 (2015).

34    Alibart, F., Gao, L., Hoskins, B. D. & Strukov, D. B. High precision tuning of state for memristive devices by adaptable variation-tolerant algorithm. *Nanotechnology* **23**, 075201 (2012).

35    Prodromakis, T., Peh, B. P., Papavassiliou, C. & Toumazou, C. A versatile memristor model with nonlinear dopant kinetics. *IEEE Trans. Electron Devices* **58**, 3099-3105 (2011).

36    Van Rossum, M. C., Bi, G. Q. & Turrigiano, G. G. Stable Hebbian learning from spike timing-dependent plasticity. *Journal of Neuroscience* **20**, 8812-8821 (2000).

37    Rubin, J., Lee, D. D. & Sompolinsky, H. Equilibrium properties of temporally asymmetric Hebbian plasticity. *Physical Review Letters* **86**, 364 (2001).

38    Debanne, D., Gähwiler, B. H. & Thompson, S. M. Heterogeneity of synaptic plasticity at unitary CA3–CA1 and CA3–CA3 connections in rat hippocampal slice cultures. *Journal of Neuroscience* **19**, 10664-10671 (1999).

**Figure 1. Metal-oxide memristive devices.** (a) SEM image of the active area of a memristive crossbar, with the particular inputs used in the STDP experiment shown on the margins. (b) Typical *I-V* curve of a memristor after its forming, with the dashed lines indicating the effective set and reset thresholds.

**Figure 2. Experimental results for spike-time-dependent plasticity.** Implemented STDP windows similar to those typical for biological synapses: in Layer 5 (the left column) and Layer 4 (the middle column) of the neocortex, and in GABAergic synapses (the right column). (a-c) The used shapes of pre-synaptic ("pre", black lines) and post-synaptic ("post", red lines) voltage pulses. (d-f) The time maxima and minima of the net voltage applied to the memristor, as functions of the time interval $\Delta t$ between the pre- and post-synaptic pulses. (g-i) The experimentally measured STDP windows, i.e. the changes of memristor's conductance as functions of $\Delta t$. The red points and black error bars show, respectively, the averages and the standard deviations of the results over 10 experiments for each value of $\Delta t$. In these experiments, the initial memristor conductance $G_0$ was always close to 33 μS.

**Figure 3**. **Modeling spike-time-dependent plasticity.** (a) The experimentally measured STDP window function (for the waveforms shown in Fig. 2a) for several initial values $G_0 = 25, 50, 75$ and 100 μS, and (b) the results of its fitting with Eqs. (1)-(3). (R-squared = 0.956.) On panel (a), lines are just guides for the eye. The inset table in panel (b) shows the used fitting parameters.

**Figure 4. Self-adaptation of spike-time-dependent plasticity.** Simulation of memristor self-tuning in a simple spiking network, using Eqs. (1)-(3) for STDP description. (a) The simulated network; (b) its equivalent circuit; (c) typical input and output spiking activity; and (d) the initial and final distributions of conductances, averaged over 10 runs, for several values of the initial conductance $G_0$. On panel (c), the top graph uses grey color coding to shows the spike initiation times. On panel (d), three middle figures show the final distribution of conductances for three values of $G_0$, after 60 s simulation. The bottom figure of panel (d) shows the final weight distribution for the hypothetical "additive" STDP model, obtained by artificially setting $\Lambda_G = 1$. The neuron parameters are as follows: $R = 4$ kΩ, $C = 1$ μF, activation threshold $U_t = 0.5$ V.

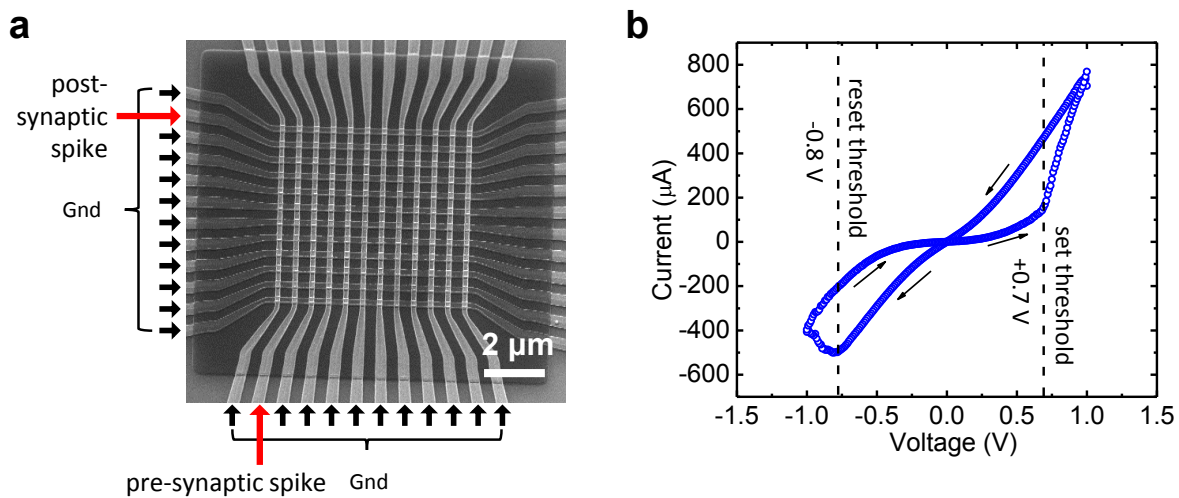

**Figure 1.**

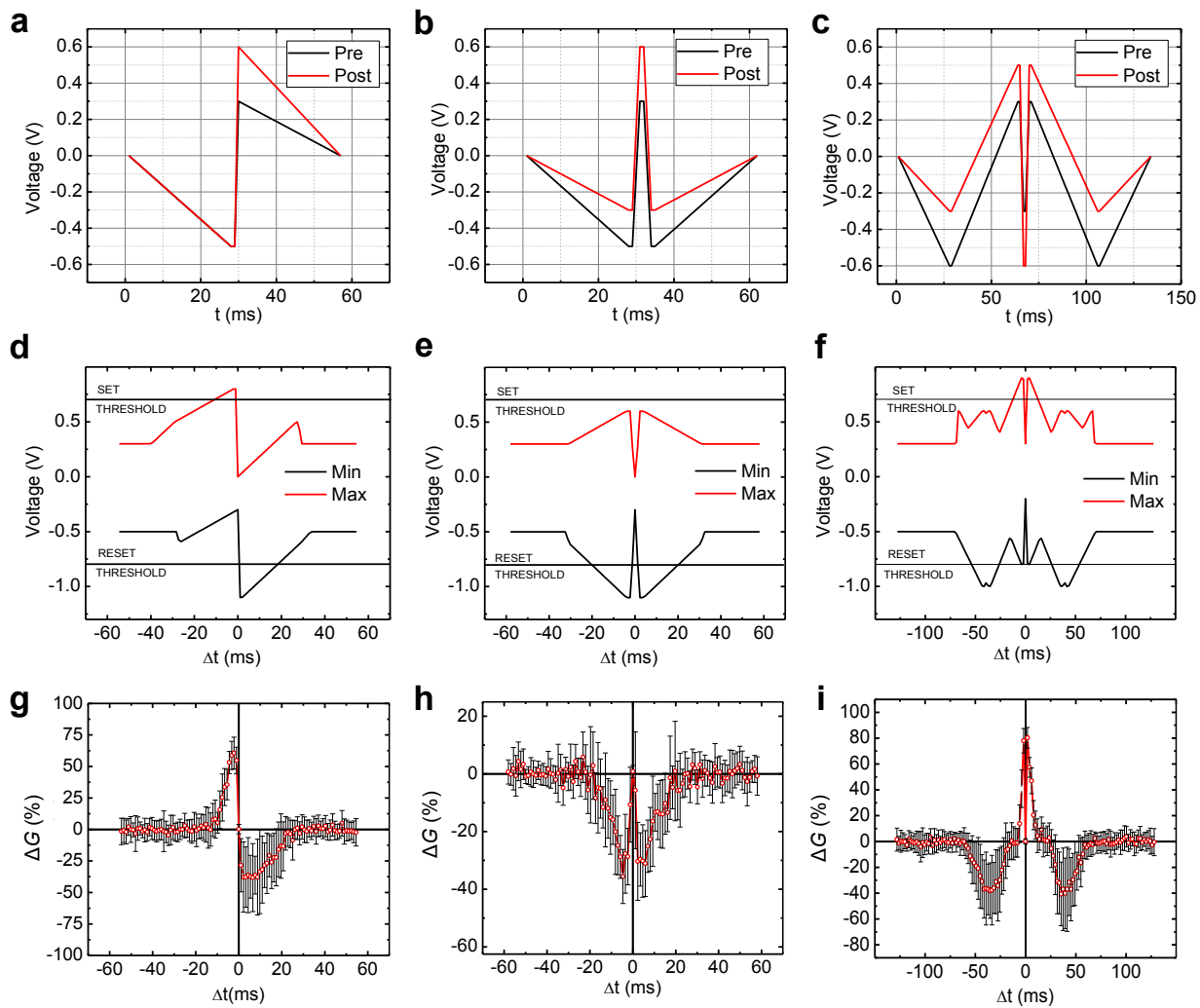

**Figure 2.**

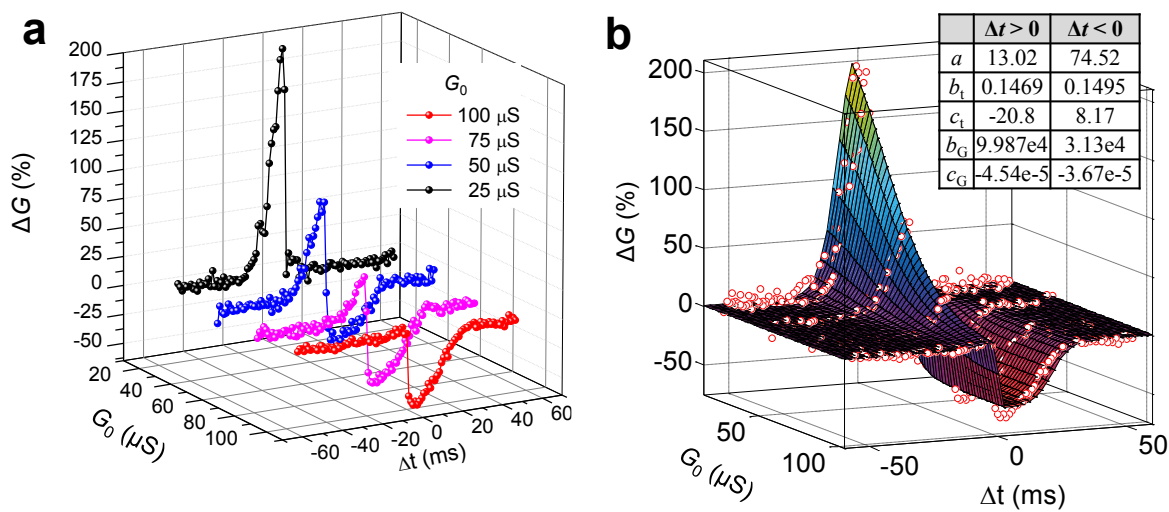

**Figure 3.**

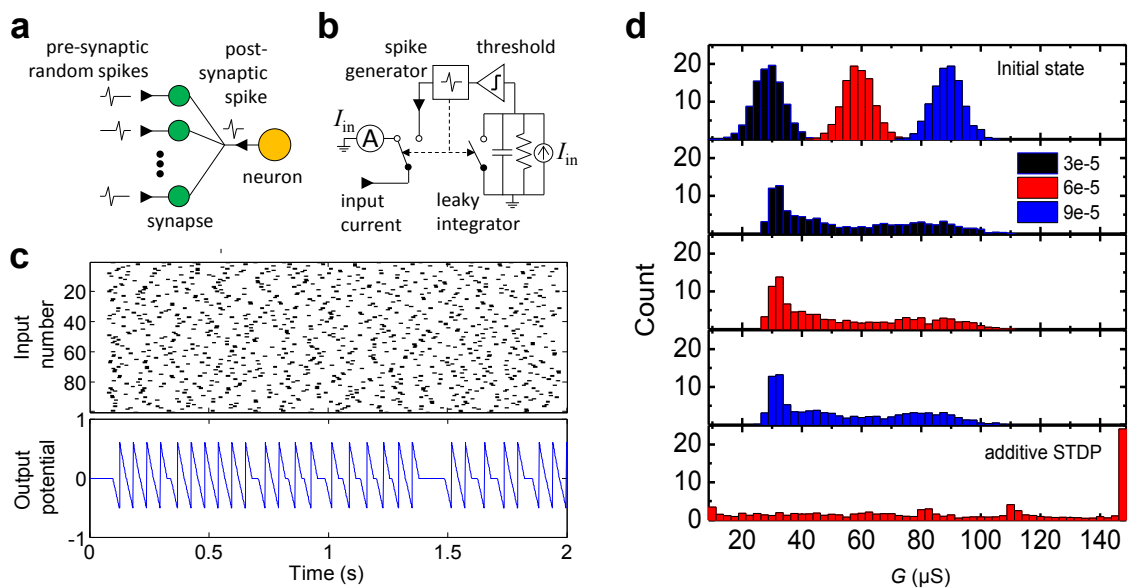

**Figure 4.**

# Supplementary Information

## Methods

The device fabrication steps were similar to those described in Ref. 1. Specifically, first, crossbar lines, 200 nm wide and separated by 400 nm gaps were formed on 4" silicon wafers covered by 200 nm of thermal $SiO_2$. After standard cleaning and rinse, fabrication started with an e-beam evaporation of Ta (5 nm)/Pt (60 nm) bilayer over a patterned photoresist to form the bottom wires. After liftoff, the wafer was descum by active oxygen dry etching at 200°C for 10 minutes. Then, a blanket film consisting of a 5-nm sputtered $Al_2O_3$ barrier and a 30-nm $TiO_2$ switching layer was deposited. This film was then removed by etching in an ICP chamber using $CHF_3$ plasma, while preserving it in the future crossbar area by pre-deposited negative photoresist. After stripping the photoresist in the 1165 solvent for 3 hours at 80°C, the wafer was cleaned using a mild descum procedure performed in a RIE chamber for 15 seconds with 10 mTorr oxygen plasma at 300 W. Next, the top electrode, consisting of 15 nm Ti and 60 nm Pt layers, was deposited by e-beam evaporation; then top wires were patterned by liftoff process. Finally, the wire bonding pads were formed by e-beam deposition of Cr (10 nm) /Ni (30 nm) /Au (500nm). All lithographic steps were performed using a DUV stepper using a 248 nm laser. After fabrication and dicing, the dies were annealed in a reducing atmosphere (10% $H_2$, 90% $N_2$) for 15 minutes at 300°C and wire-bonded to a DIP40 package. The final crossbar layout is shown in Fig.1a.

All electrical characterizations were performed using the Agilent B1500A parameter analyzer with Agilent B1530 arbitrary waveform generator modules. In addition, the Agilent B5250A switching matrix was employed for selecting one device from the crossbar and carrying out the experiment. Before performing the STDP experiments with a memristor, it was electrically formed by applying 210 µA, reaching 2.6 V in a current-controlled sweep. After that, the device switching thresholds were measured by applying a double $I\text{-}V$ ramp (Fig. 1b); they have turned out to be close to 0.7 V for the set operation (i.e. switching from low to high conductive state), and close to -0.8 V reset operation, i.e. for the reset operation.

## Self-Tuning Dynamics

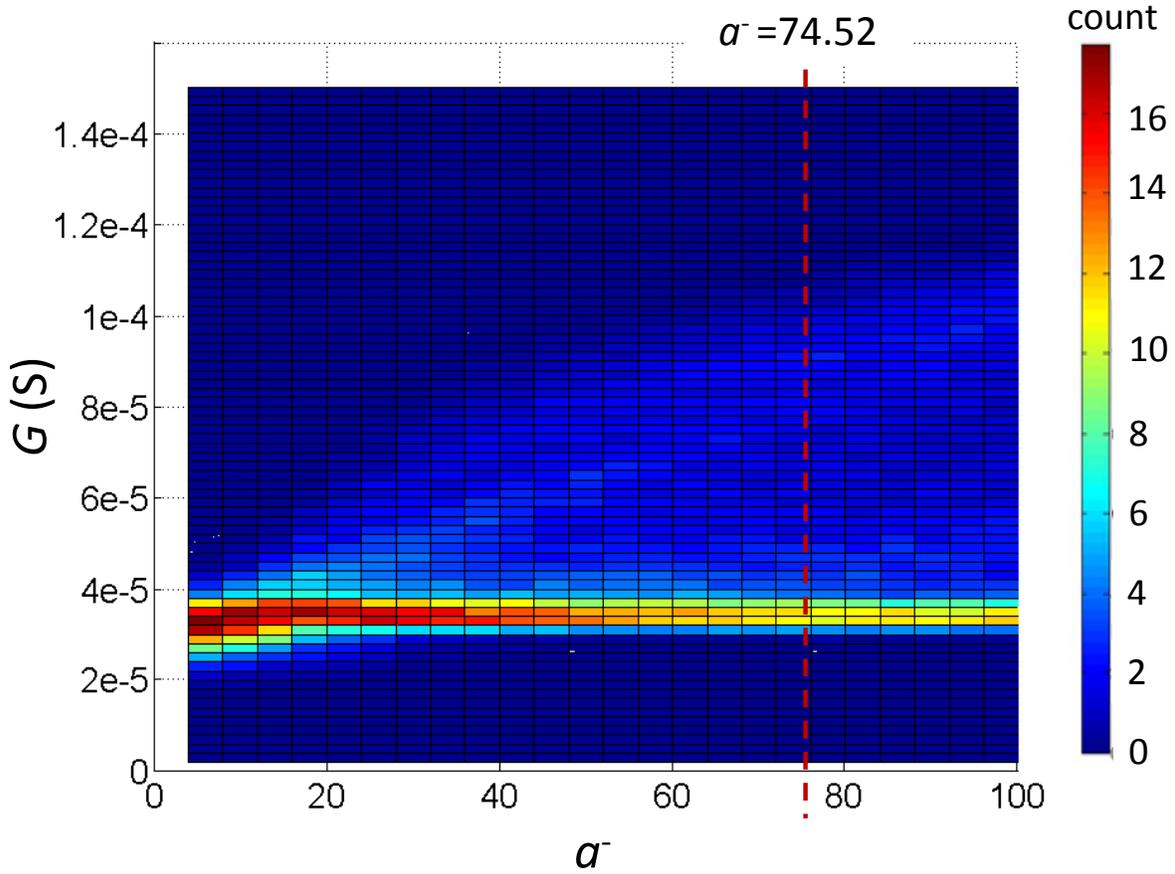

**Figure S1.** Map of the final distribution of memristor conductances, obtained in the spiking network simulation, using the STDP model described by Eq. (1) – (3), for various choices of the fitting constant $a^-$. In particular, the plot shown in the middle panel of Fig. 4d is a vertical slice of this map for the particular value $a^- = 74.52$, shown with the vertical dashed line. The values of other fitting constants and model assumptions are the same as shown in the inset of Fig. 3b and Fig. 4 of the main text, respectively.

1    Prezioso, M. *et al.* Training and operation of an integrated neuromorphic network based on metal-oxide memristors. *Nature* **521**, 61-64 (2015).